# Are Game Platforms suitable for Parkinson Disease patients?

Ioannis Pachoulakis[1], Nikolaos Papadopoulos[1] and Cleanthe Spanaki[2]

[1]Department of Informatics Engineering
Technological Educational Institute of Crete
Heraklion, Crete, Greece
*{ip@ie.teicrete.gr,mpt24@edu.teicrete.gr}*

[2]Department of Neurology
Faculty of Medicine
University of Crete
Heraklion, Crete, Greece
*{kliospanaki@med.uoc.gr}*

**Abstract:** Parkinson's Disease (PD) is a progressive neurodegenerative movement disorder that affects more that 6 million people worldwide. Motor dysfunction gradually increases as the disease progress. It is usually mild in the early stages of the disease but it relentlessly progresses to a severe or very severe disability that is characterized by increasing degrees of bradykinesia, hypokinesia, muscle rigidity, loss of postural reflexes and balance control as well as freezing of gait. In addition to a line of treatment based on dopaminergic PD-specific drugs, attending neurologists strongly recommend regular exercise combined with physiotherapy. However, the routine of traditional rehabilitation often create boredom and loss of interest. Opportunities to liven up a daily exercise schedule may well take the form of character-based virtual reality games which engage the player to physically train in a non-linear and looser fashion, providing an experience that varies from one game loop the next. Such "exergames", a word that results from the amalgamation of the words "exercise" and "game" challenge patients into performing movements of varying complexity in a playful and immersive virtual environment. In fact, today's game consoles using controllers like Nintendo's Wii, Sony PlayStation Eye and the Microsoft Kinect sensor present new opportunities to infuse motivation and variety to an otherwise mundane physiotherapy routine. But are these controllers and the games built for them appropriate for PD patients? In this paper we present some of these approaches and discuss their suitability for these patients mainly on the basis of demands made on balance, agility and gesture precision.

## 1. INTRODUCTION

Parkinson's disease (PD) is a neurodegenerative condition which affects parts of the brain that control body movement. More precisely, PD results from the loss of dopamine-producing neurons in a mid-brain region known as "substantia nigra", responsible for - among others - the smooth and purposeful coordination of body muscles. It is still unclear why these cells lose function. According to the European Parkinson's Disease Association (EDPA) [1], around 6.3 million people suffer from the disease worldwide. Symptoms progress slowly but irreversibly, so that late stages are more abundant in the elderly (>60 year old) population, with only approximately 10% of PD patients under the age of 50. In early stages, PD commonly affects motor function, while cognitive, behavioral and mental-related symptoms are usually met in more advanced stages [2]. Non motor-related symptoms of the disease include sleep disturbance, depression, anxiety, psychosis, visual hallucinations, cognitive impairment, pain and fatigue. The four distinct, fundamental motor symptoms of the disease are grouped under the acronym TRAP: Tremor, Rigidity, Akinesia (or bradykinesia) and Postural instability [3]. These symptoms, that often develop in various combinations, hinder common daily activities and trouble patients' social relationships thus reducing the quality of life, especially as the disease progresses [4],[5].

Although a cure has not yet been discovered for PD, medications usually help control the symptoms and maintain body functionality at reasonable levels throughout the lifetime of the patient. According to the American Parkinson Disease Association (APDA) [6], six categories of drugs are proposed for PD condition therapy: levodopa, dopamine agonists, MAO-B inhibitors, COMT inhibitor, anticholinergic agents and amantadine. Standard medication paths cannot be easily achieved, as the course and symptoms of the disease vary significantly among patients [7]. Based on current clinical practice, levodopa is considered as the most efficient drug for improving PD-related motor symptoms, although large doses over an extended period of time have been connected to the development of involuntary abnormal movements called dyskinesias that further aggravate the patients' walking ability and motor function. Recent common practice starts patients on agonists, postponing levodopa for later stages when motor symptoms are not satisfactorily controlled [8]. In more advanced stages, combinations of levodopa, dopamine agonists, COMT inhibitors and MAO-B may be used to control symptoms and achieve optimal results [7].

Adding to the value of medical treatment, physiotherapy appears highly effective in controlling PD-related symptoms. A number of Parkinson clinical facilities and associations provide physical activity guidelines, suggesting daily activities and tasks, even diet schedules. For example the Parkinson Society of Canada [9] provides online detailed instructions on how to correctly perform stretching and other physical exercises. Exercise interventions in randomized controlled trials [10] prove that physical exercise such as stretching, aerobics, unweighted or weighed treadmill and strength training improves motor functionality (leg stretching, muscle strength, balance and walking) and quality of life. In addition, the "training BIG" strategy for PD rehabilitation [11] has also shown promising results: exercises that focus on amplitude training lead to faster upper and lower limb movements. Training BIG prescribes exercises at maximum range of motion (maximum amplitude) which deploy the entire body of the patient both in seated and standing posture such as reaching and twisting to each side or stepping and reaching forward. However, traditional physiotherapy for PD can easily bore and tire patients due to repetition. Opportunities to spice up the daily exercise schedule can come from today's game consoles and supported virtual reality exergames which by design combine exercises in a non-linear and loose game context. The following section presents recent approaches using Nintendo's Wii, Sony PlayStation Eye and the Microsoft Kinect sensor.

## 2. RELATED WORK

Yu et al [12] developed a real-time multimedia environment based on a 10 near-IR camera Motion Analysis System to capture a patient's body in 3D using attached retro-reflecting markers. Guided by visual and auditory cueing, patients executed exercises based on the BIG strategy [11]. Incoming data streams were combined and mapped on an on-screen avatar that replicated the patient's movements (Figure 1). The solution, however, is not cost effective due to the need of installing multiple IR cameras and calibrating the system. In addition, it is hard for people with PD condition to use the system on their own, without help (markers must be attached to their body at specific points every time they need to exercise).

In the context of the EU-funded CuPiD project, Tous et al. extended the tele-rehabilitation platform Play For Health (P4H) to create exergames for PD patients [13]. Three games were created: Touch 'n' Explode (pop objects in the virtual environment), Stepping Tiles (step on kitchen tiles) and Up 'n' Down (sit - stand exercise). During the game, body movement is translated to 3D Avatar movements in a virtual environment using data collected from a Body Area Network of wearable sensors developed specifically for the project. In addition, the system was able to capture freezing motion incidents.

The Nintendo Wii gaming system has also been evaluated as a possible rehabilitation tool for PD. The Wii system uses a handheld controller to communicate with a console installed by the display monitor / TV. Data such as the rotation and acceleration of the controller are sent to the console wirelessly. In addition, the "balance board", a rather common extender component used by several Wii games contains several sensors which calculate the mass of the player and his/her center of gravity. Case studies such [14] have evaluated Wii as an off-the-self, cheap solution that already provides game titles in the area of rehabilitation. Data related to the improvement of gait, balance, cognitive reaction and impaired functional mobility were recorded and presented. In most cases the Wii Balance board extender was used, as it is required by a wide variety of Wii games.

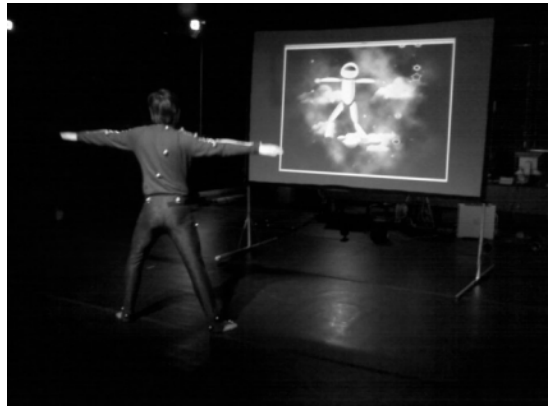

**Figure 1: An interactive multimedia system for PD rehabilitation (Yu et al)**

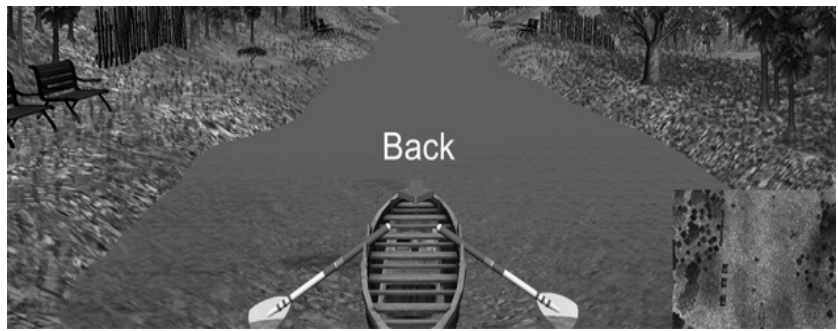

**Figure 2: Wii-based game for Parkinson rehabilitation (Paraskevopoulos et al)**

Two Wii-based rehabilitation games were also developed by Paraskevopoulos et al [15] using PD-specific movements extracted from literature. In the first game the patient controls a two-paddle row boat from a seated position to reach a specific point in a certain amount of time (Figure 2). In the second game (water valve mini golf game) the player rotates a valve several times to guide a ball through a pipe into a hole. Data streamed from the sensor's accelerometer and gyroscope were combined to provide improved orientation and linear motion results so as to map the user's movements onto a 3D avatar. These game exercises are intended to improve speed, rigidity, range of movement and bilateral mobility.

Evaluating the advantages/or disadvantages of similar (to the Wii) technologies, Assad et al [16] conclude that PD patients are not able to coordinate effectively by moving the handheld controller and pushing buttons on it at the same time. So, they examined the Sony PlayStation Eye system as a potential rehabilitation tool for the PD condition. They developed WuppDi!, a collection of five PD-oriented motion-based games, most of which required one or two hand markers such as a glove or a wooden stick to interact with virtual game objects. Participants welcomed this approach as a physical activity albeit having trouble handling the input devices.

Evidence so far weighs on the side of a consensus that PD patients are to benefit mostly from exergame systems that accurately capture full body motion without the need of external handheld or wearable devices. In their review, Gillian et al [14] propose that platforms such as the XBOX Kinect which do not require raised platforms, handheld controllers and/or body markers need to be evaluated with respect to rehabilitation opportunities with maximal safety for PD patients. In addition, such systems are natural candidates for home based solutions that can be easily used by PD patients without lengthy guided instructions on connecting peripherals. Microsoft Kinect requires no external input controllers and captures the motion of the entire human body in 3D using a single sensor bar in front of the player. Kinect-based exergames tailored to PD are however sparse due to the fact that the technology is relatively new. The first generation of Kinect sensors was released in November of 2010 and was paired to the Xbox 360 console, while in June 2011 Microsoft released a Software Development Kit (SDK) to allow the creation of Kinect-based applications.

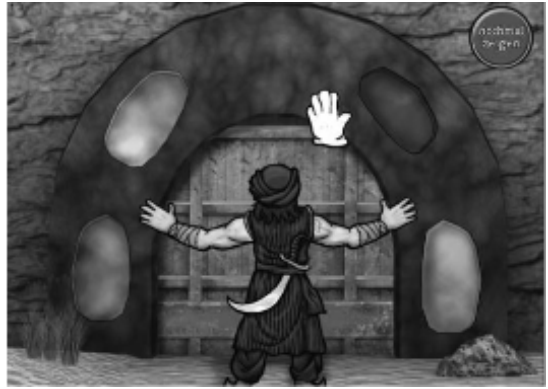

**Figure 3: Games based on PlayStation Eye (Assad et al)**

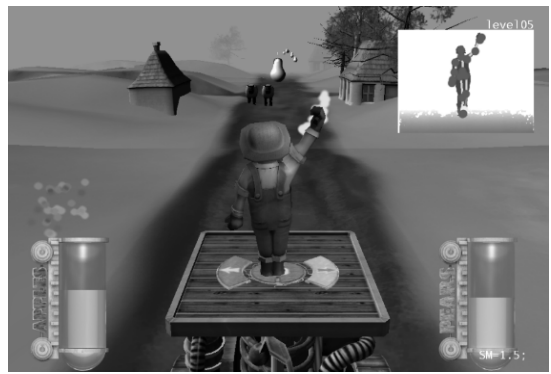

**Figure 4: Kinect-based game (Galna et al)**

Galna et al [17] developed a Kinect game for PD patients which includes upper and lower limb movements to improve dynamic postural control. The upper torso of the player is mapped onto a farmer avatar who drives a tractor, collects fruits and avoids obstacles in a 3D environment. Fruit collecting is achieved by specific hand movements of the patient, choosing to move to the right or to the left hand depending on the color of the fruit. Large steps (front, back and sideways) with one foot centered in a stationary position guide the tractor to avoid obstacles like sheep, high wires, birds, etc. To maintain patient motivation, the game has levels of increasing difficulty and game complexity which range from simple hand movements to more complex activities requiring cognitive decisions combined to physical movement and dual tasking (simultaneous hand and foot work). The design principles for the game resulted from a workshop where participants played and evaluated a wide range of commercial games developed for Microsoft Xbox Kinect and Nintento Wii. The workshop once more revealed the difficulty some patients have to interact with Wii's handheld controller and Balance Board. A pilot test of the game led to several useful conclusions on gameplay design, feasibility and safety. The authors conclude that Kinect-based games seem both safe and feasible for PD patients, although the use of Kinect as a home-based rehabilitation solution needs further investigation.

Another multiplayer game also based on the Kinect sensor was developed by Hermann et al [18] to investigate whether game cooperation improves communication and coordination. Using hand movements, participants had to collect buckets of water in a flooded area to reveal an object underneath. The game was played in two different modes. In the first mode (loose cooperation) both players drained the area in parallel, while in the second mode (strong cooperation) only one would drain, while the second player's duty was to reveal the object. Gameplay was recorded to reveal information regarding the discussion and level of cooperation between the participating patients. Subsequent analysis showed that multiplayer games are feasible for this kind of target group and that asymmetric roles (strong cooperation) can motivate communication between the participants and lead to a better game experience.

3. DISCUSSION

Parkinson's disease patients following a long-term repetitive exercise schedule can easily get bored, lose interest and eventually drop out of a rehabilitation program. Exergames on the other hand engage patients into repeatedly executing simple or complex exercise patterns within a goal-oriented enjoyable context, with real-time feedback and rewards. Successful exergames should also draw from existing recommended PD training programs.

However, not all game platforms, hardware controllers and games seem appropriate for PD patients. For example, lack of postural stability that characterizes the PD condition may lead to falls and severe injuries, so that raised platforms such as Wii's Balance board should be avoided. In addition, PD patients wearing gloves or using handheld controllers with buttons encounter difficulties interacting with a virtual game environment. It may seem plausible that patient safety is maximally preserved with gaming systems which provide full body motion tracking without additional / external add-ons that users must interact with. Low cost solutions with minimal footprint that can be easily set up either in a clinical setting or at a home environment would naturally be preferred. Still, all such systems must be evaluated with a keen eye on patient safety.